\documentclass[a4paper,fleqn]{cas-dc}
\usepackage[authoryear,longnamesfirst]{natbib}
\usepackage{graphicx}
\usepackage{xcolor}
\usepackage{ulem}
\usepackage{hyperref}
\usepackage{amssymb}
\def\tsc#1{\csdef{#1}{\textsc{\lowercase{#1}}\xspace}}
\tsc{WGM}
\tsc{QE}

\ExplSyntaxOn
\keys_set:nn { stm / mktitle } { nologo }
\ExplSyntaxOff

\begin{document}
\let\WriteBookmarks\relax
\def\floatpagepagefraction{1}
\def\textpagefraction{.001}

\shorttitle{Canonical treatment of strangeness and light nuclei production}
\shortauthors{N.~Sharma}  
\title [mode = title]{Canonical treatment of strangeness and 
light nuclei production}
\author{Natasha Sharma}[orcid=0000-0001-8046-1752]
\ead{nsharma@cern.ch}
\affiliation{organization={Indian Institute of Science Education and Research (IISER) Berhampur},
            addressline={Ganjam}, 
            city={Odisha},
            postcode={760003}, 
            country={India}}

\begin{abstract}
The canonical effects on strangeness and light nuclei production in high energy collisions are investigated, with a particular focus on low multiplicity events observed in small collision systems  at the Large Hadron Collider (LHC), and also in 
low energy collisions at the Relativistic Heavy Ion Collider (RHIC). We analyzed the yields of various particles, such as pions, protons, lambdas, multi-strange hadrons and light nuclei, and their dependencies on the produced charged particle multiplicities using the Statistical Hadronization Model (SHM). Our findings indicate that the canonical treatment of strange and baryon quantum numbers are essential for describing the particle production 
in these collisions, particularly in small collision systems.
\end{abstract}

\maketitle

\section{Introduction}
The study of multi-strange hadrons and light nuclei production in high energy collisions provide insights into the understanding of particle production mechanism in the high energy collisions. 
It is observed experimentally that the production of multi-strange and light nuclei depend on the total number of produced charged particle in these collisions.  
These dependencies can be related to the canonical effect  in these collisions. 
This work aims to elucidate that the  canonical treatment of strangeness and baryon quantum numbers are essential for understanding the dynamics of small systems at the LHC energies and heavy-ion collisions at the lowest RHIC energy. 


The Statistical Hadronization Model (SHM) has been very successful in explaining the light flavour particle production including the light nuclei and hypernuclei measured by A Large Ion Collider Experiment (ALICE) in Pb-Pb collisions at $\sqrt{s_{\rm NN}}$ = 2.76 TeV at the LHC. 
The protons to pions ratio is found to remain almost constant with the produced charged particle multiplicity in the Pb-Pb collisions. This is related to the common chemical freeze-out temperature ($T_{ch}$) = 156.5 MeV. However, the suppression of the strange particle to pion yield ratios with the decreasing produced charged particle multiplicity ($dN_{ch}/d\eta$)  
is observed in the pp and p-Pb collisions i.e. in the small collision systems \cite{ALICE:2017jyt}. The
suppression effect on ratios with decreasing multiplicity is found to be more for particles with more strangeness content (|S|). This effect can be linked to the “strangeness canonical effect”. 

In the  Grand Canonical (GC) ensemble, the quantum numbers are conserved on  average and  are  implemented  using the corresponding  chemical potentials
$\vec\mu$ linked to conserved charges $\vec Q$.
The partition function depends on thermodynamic quantities and the Hamiltonian describing the system, 
\begin{equation}
\label{eq1}
Z_{GC} = \textrm{Tr} \left[ e^{-(H-\vec \mu\cdot \vec Q)/T}\right]. 
\end{equation}

\section{Strangeness suppression in small collision systems at the LHC and lower RHIC energies}

The strangeness suppression applies particularly for more peripheral collisions or lower energies, where the canonical formulation of thermodynamics is applicable.

At LHC energies, the particles and anti-particles are produced in almost equal abundances, which leads to $\vec \mu=0$. In case of exact strangeness conservation and assuming that all other quantum numbers are conserved on average in the GC ensemble. 
This is a good approximation for the thermal description of the collision fireball produced at the LHC energies.  
In this case, the canonical ensemble with exact implementation of strangeness conservation and total strangeness S = 0
is  achieved by introducing a  delta function under the trace in Eq.~\eqref{eq1},

\begin{equation}
Z^C_{S=0} = \textrm{Tr}\left[ e^{-H/T}\delta_{(S,0)}\right]  .
\end{equation}

The multiplicity of a particle with strangeness quantum number $S$ in the given experimental acceptance is given as, 
\begin{align}
\label{equ10}
\langle N_k^s\rangle_A \simeq V_A \, n_k^s(T) \, \frac{I_{s}(S_1)}{I_{0}(S_1)}.
\end{align}
where, $V_A$ is the effective fireball volume and ${I_{s}(S_1)/ I_{0}(S_1)}$ ratio is the suppression factor which  decreases in magnitude   with increasing  $s$ of  hadrons  and with decreasing thermal phase-space occupied by strange particles. The argument  $S_1$  of the Bessel functions is $S_1=V_C\sum_k n(k,T)$, where $V_C$ is the full space volume where $S$ is exactly conserved

In the following, we apply the HRG model with the canonical treatment of strange quantum number (SCE) described above to quantify   production of (multi-)strange hadrons and their 
behavior with charged particle multiplicity  as observed by the ALICE collaboration in different colliding systems and collision energies at the LHC.  We also correct the HRG model taking into account interactions among hadrons using S-matrix corrections based on
known phase shift analyses`\cite{Cleymans:2020fsc}.

We performed the thermal model fits to the data collected by the ALICE collaboration for pp collisions at $\sqrt{s}$ = 7 TeV~\cite{ALICE:2017jyt} as well as 
in  p-Pb collisions at $\sqrt{s_{\rm NN}}$ = 5.02 TeV~\cite{Adam:2015vsf}
for various multiplicity classes. We have used the Strange Canonical ensemble approach as implemented in the THERMUS package~\cite{Wheaton:2004qb} which is further extended to include hadron interactions within the S-matrix approach.
We have fixed the temperature value to 156.5 MeV as 
given by the LQCD calculations and strangeness suppression factor ($\gamma_{s}$) is fixed to 1 in order to quantify the effects of fitting on the correlation volume ($V_C$) and on the volume of system in the experimental acceptance ($V_A$). 
We found that for large multiplicities  $V_A \simeq  V_C$, 
therefore  we first put them equal for all multiplicities.

\begin{figure*}[ht] 
\center
\includegraphics[scale=0.55]{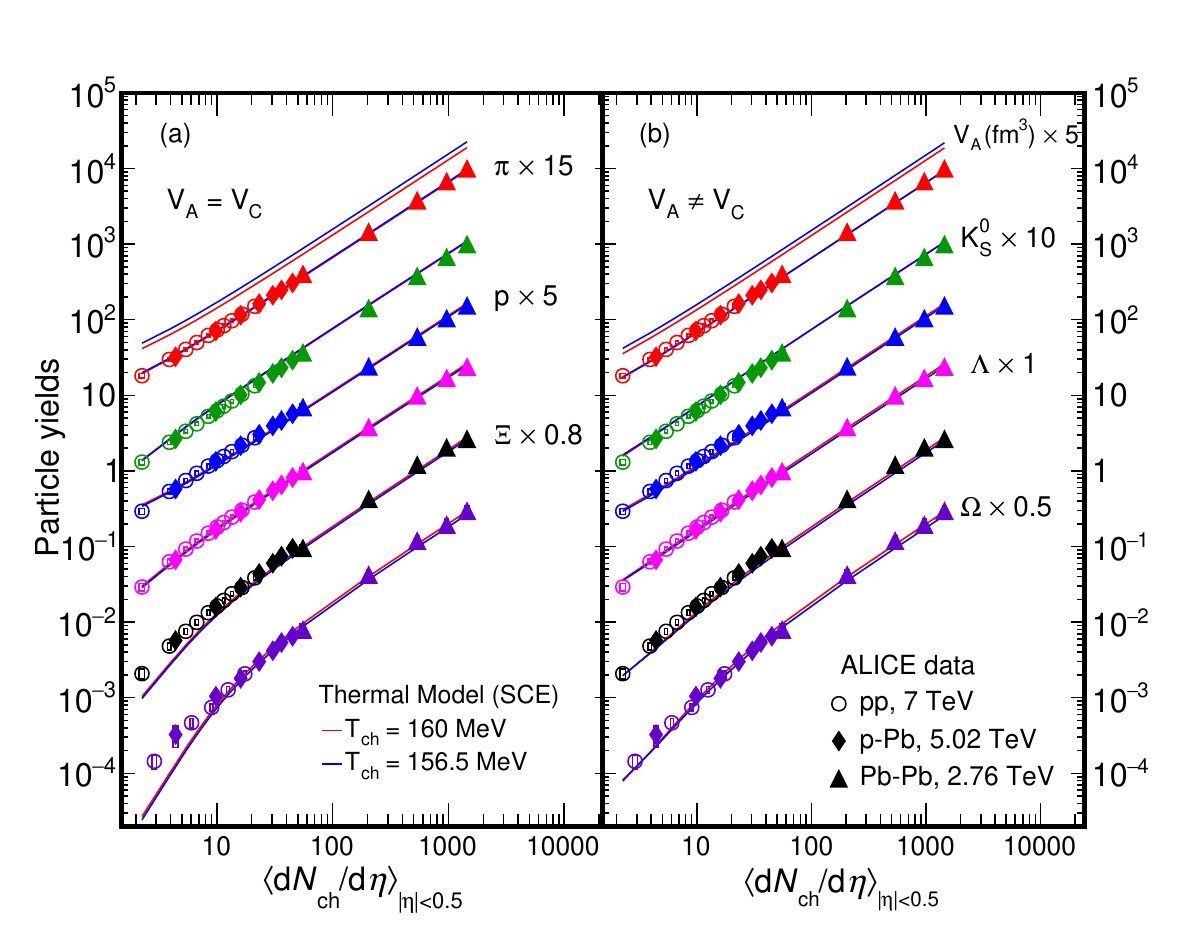}
\vskip -0.2 cm
\caption {
Left: Yields for $V_A=V_C$. Right: Yields for $V_A\neq V_C$,
 The particle yields are indicated by symbols, different symbols are used for different collision systems. The lines represent  the SHM model results. The solid blue lines have been calculated for $T$ = 156.5 MeV while the solid red lines have been calculated for $T$ = 160 MeV. }  
\label{fig:fig5}
\end{figure*}

Figure  \ref{fig:fig5} (left) shows the yields of hadrons calculated in the SCE for different charged particle multiplicities
$dN_{\rm ch}/d\eta$ at fixed temperature and for a single volume  $V=V_A\simeq
V_C$. 
The use of single volume  leads to too much suppression at small charged particle multiplicities,  particularly for $S=-2$ and $S=-3$
baryons. 
As the strangeness conservation is related to the full phase-space and  particle yields are measured in some acceptance window, thus,  $V_C$ can be larger than $V_A$. We have performed the SCE model fit  to data  with two independent volume parameters as shown in Fig. \ref{fig:fig5} (right). The resulting yields  exhibit  much better agreement with data by decreasing strangeness suppression at lower multiplicities due to  larger value of $V_C$ than $V_A$.  

We have fitting the particle yields for different multiplicity classes for pp collisions at 7 TeV. The obtained $V_A$ and $V_C$  for each multiplicity bins are fitted with linear functions of charged particle multiplicity. The obtained parametrization for $T_{ch}$ = 156.5 TeV are:

\begin{align}
    V_A &= 1.55 + 3.02 \times \frac{dN_{ch}}{d\eta} \\
    V_C &= 12.32 + 3.02 \times \frac{dN_{ch}}{d\eta}
\label{eq:volume}
\end{align}

These parametrization are used to predict particle yields for p-Pb collisions at  $\sqrt{s_{\rm NN}}$ = 5.02 TeV as well as for Pb-Pb collisions at  $\sqrt{s_{\rm NN}}$ = 2.76 TeV and are  shown in Fig.  \ref{fig:fig5}. The experimental particle yields are indicated by symbols, different symbols are used for different collision systems. The lines represent  the SHM model results.  The experimental data is found to be consistent with the SHM predictions over a wide  range of multiplicity.

 
  In the GC ensemble with $\vec \mu=\vec{0}$ and for constant temperature the yields of all particles  per volume is independent of $dN_{ch}/d\eta$. On the other hand, the density of charged particles in the SCE model exhibits a non-linear dependence on volume parameters.  
 Indeed, in the context of strangeness production at the LHC,  already   from a previous analysis~\cite{Sharma:2018jqf} it was clear that the canonical ensemble with exact strangeness conservation (CSE)  is the best case  for describing particles yield with non-zero strangeness quantum number in the low multiplicity events or in small (pp ad p-Pb) collision systems at the LHC energies. 


Recently, the $K^0_S$ and $\Lambda$ yields have been experimentally measured by the STAR collaboration for Au-Au collisions at $\sqrt{s_{\rm NN}}$ = 3 GeV~\cite{STAR:2024znc}. 
The dependence of particle ratios on the collision energy shows that the strangeness production is suppressed in the heavy-ion collisions at lower RHIC energies (i.e. $\sqrt{s_{\rm NN}} < $ 5 GeV). The comparison of the experimentally measured particle ratios with the thermal model predictions are reported in Fig. 8 of the Ref.~\cite{STAR:2024znc}. It is seen that for the collision energies greater than 5 GeV, the thermal model predictions with the grand canonical (GC) ensemble and with the strange canonical (SC) ensemble are consistent with the experimental data. However for the  lower RHIC energies, the GC ensemble over predicts  the particle ratios and the SC ensemble with the correlation radius ($R_C$) between 2.9 fm to 3.9 fm describes the particle ratios well.


We conclude that the strangeness production is suppressed in the small collision systems at the LHC energies and in the heavy-ion collisions system at lower RHIC energies and hence, the strangeness quantum number has to be treated canonically to describe the yields of strange particle in such collisions.

\section{Baryon suppression in small collision systems and at lower RHIC energies}

\begin{figure}
\includegraphics[scale=0.45]{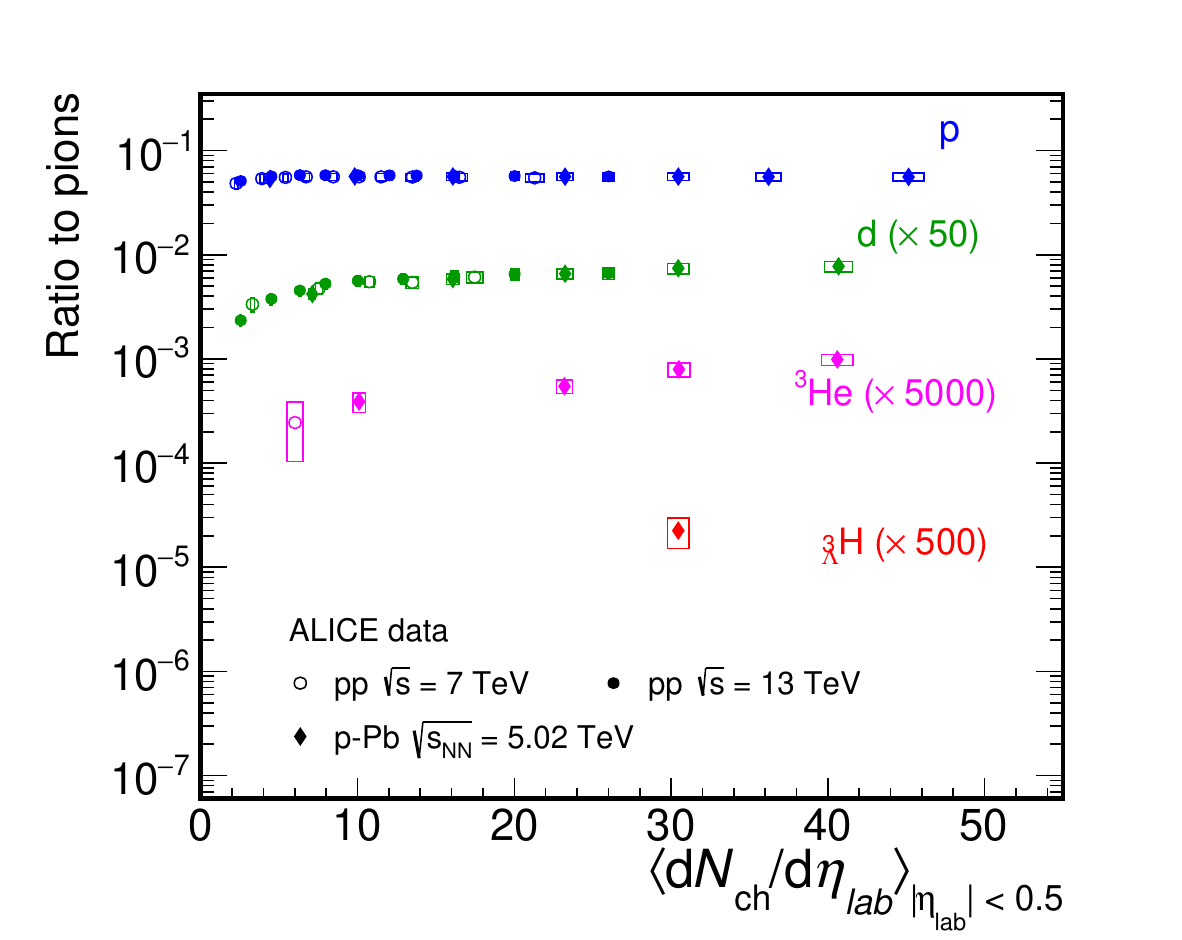}
\caption{Ratio of light nuclei to pions yields as a function of charged particle multiplicity measured by the ALICE collaboration in pp, p-Pb and Pb-Pb collisions at the LHC energies.}
\label{fig:data2pions}
\end{figure}

The production of light nuclei such as deuterons, tritons, $\rm ^3He$ and their antiparticles has been measured by the ALICE collaboration at various charged particle multiplicity bins in pp, p-Pb and Pb-Pb collisions at the LHC energies~\cite{ALICE:2015wav,ALICE:2017xrp,ALICE:2019fee,ALICE:2019bnp,ALICE:2020foi}.
Figure \ref{fig:data2pions} shows the ratio of light nuclei to pions yields as a function of charged particle multiplicity bins as measured by the ALICE collaboration in pp, p-Pb and Pb-Pb collisions. 
The ratios are found to remain constant as a function of charged particle multiplicity in Pb-Pb collisions whereas in pp and p-Pb collisions the ratios  increase with charged particle multiplicity. The ratios in small systems merge with the ratios measured in the heavy-ion collisions for similar charged particle multiplicity. 

Hence, we conclude that in the small collision systems, the baryons to pion ratios is suppressed with decreasing multiplicity and increasing baryon quantum number of hadrons. This observation can be seen as the “baryon canonical suppression effect”.
This requires treatment of baryon quantum number canonically which is analogous to the strangeness canonical effect as discussed in the previous section.

The data on light nuclei are obtained in a kinematical region where  net quantum numbers like 
net baryon number and net strangeness are zero as witnessed by the particle-antiparticle symmetry
observed in the central rapidity region at the LHC. To take this into account we focus on a system which
has zero net baryon number inside a correlation volume $V_C$.
The partition function  is constructed 
by inserting a Kronecker $\delta$ function, thus enforcing zero baryon number:  
\begin{equation}
Z^C_{B=0} = \textrm{Tr}\left[ e^{-H/T}\delta_{(B,0)}\right]  .
\end{equation}

The multiplicity of a particle with baryon quantum number $B$ in the given experimental acceptance is given as, 
\begin{align}
\label{equ10}
\langle N_k^b\rangle_A \simeq V_A \, n_k^b(T) \, \frac{I_{b}(B_1)}{I_{0}(B_1)}.
\end{align}
where, $V_A$ is the effective fireball volume, $B$ is taken as exactly conserved quantity and ${I_{b}(B_1)/ I_{0}(B_1)}$ ratio is the suppression factor which  decreases in magnitude   with increasing baryon number ($b$) of  hadrons. The argument  $B_1$  of the Bessel functions is $B_1=V_C\sum_k n(k,T)$ where $V_C$ is the full space volume where 

We performed the thermal model fits to the $\pi$, protons (p) and deuterons (d) yields measured by the ALICE collaboration in pp collisions at $\sqrt{s}$ = 13 TeV \cite{Acharya:2019kyh}. 
for various multiplicity classes. 
We have used the Baryon Canonical ensemble (BCE) approach as implemented in the THERMUS package~\cite{Wheaton:2004qb} to perform the fitting to the data. 
As discussed in the previous section, the temperature was fixed to 156.5 TeV and $\gamma_s$ was fixed to 1. 
The light nuclei production are measured in the same acceptance volume as the strange particle production, so we have used the same acceptance volume as mentioned in Eq. 4. 
The only free parameter is the correlation volume ($V_C$).

\begin{figure}
\includegraphics[scale=0.45]{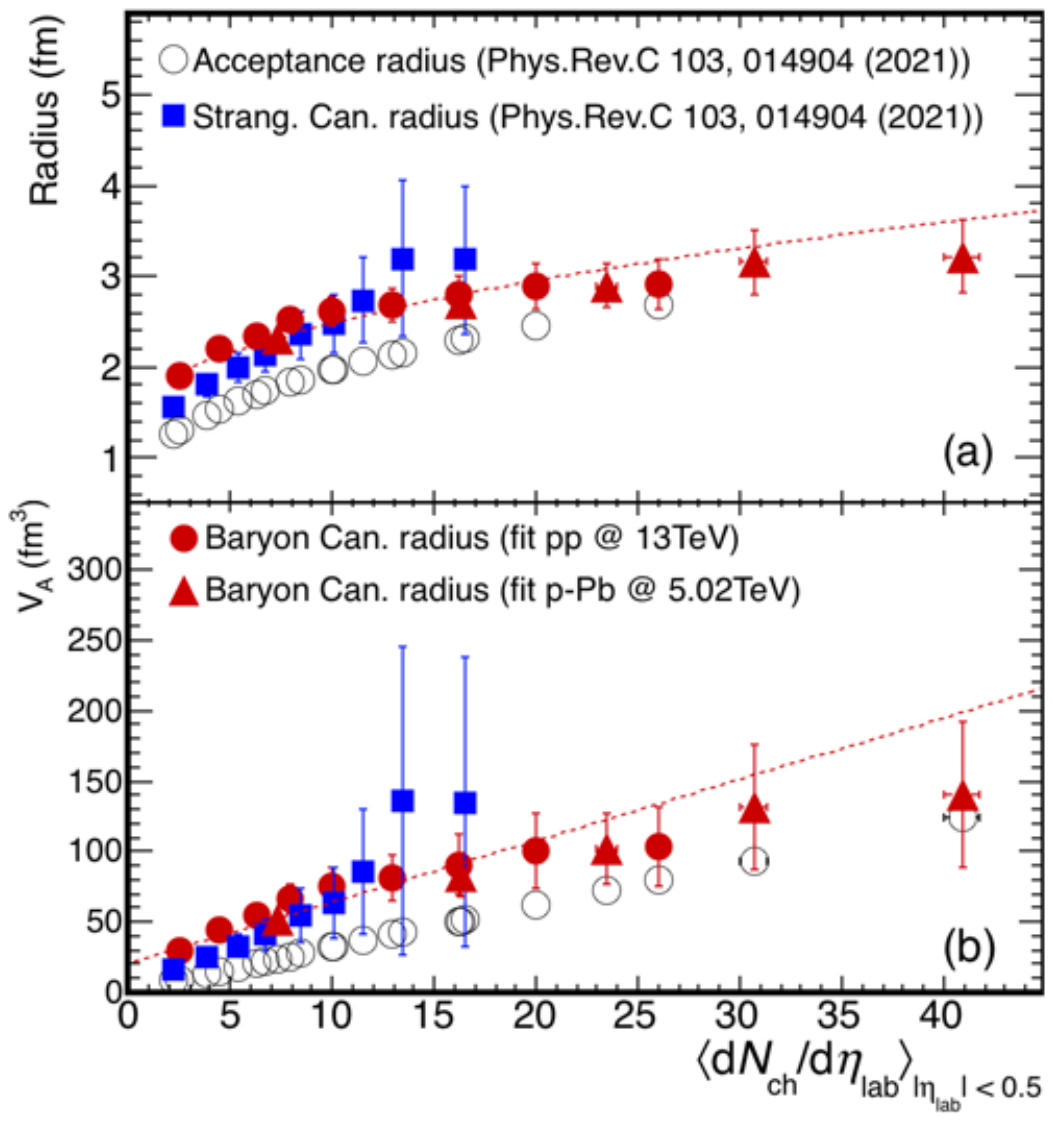}
 \vskip -2.2 cm
\caption{(a) Radii and (b) Volumes obtained from the Statistical thermal model fit to the ALICE data as a function of charged
particle multiplicities. See text for more details.}
\label{fig:radiusVolume}
\end{figure}

The top panel of Fig. \ref{fig:radiusVolume} shows the obtained radii and the bottom panel show the volumes from our analysis as a function of charged particle multiplicity for various multiplicity classes in pp and p-Pb collisions. The open circle represents the acceptance radius and volume, the solid squares represent correlation volume where strangeness is conserved exactly~\cite{Cleymans:2020fsc} while the solid circle and triangle represent correlation volume where baryon number is exactly conserved~\cite{Sharma:2022poi}.   The dotted line represents the linear fit to the baryon canonical correlation volume.
For low multiplicity events, the correlation volumes are found to be larger than the acceptance volume as correlation volume is the full space volume and can be larger than the experimental acceptance volume.  
The acceptance and correlation volume are converging to same value for the charged particle multiplicity greater than 15 i.e. the experimental acceptance volume and the correlation volumes are becoming equal for $dN_{ch}/d\eta \ge 15$. This can be interpreted as for high charged particle multiplicity, the strangeness and baryon quantum number are conserved globally and canonical treatment of these quantum numbers are not required. 

\begin{figure}
\center
\includegraphics[scale=0.40]{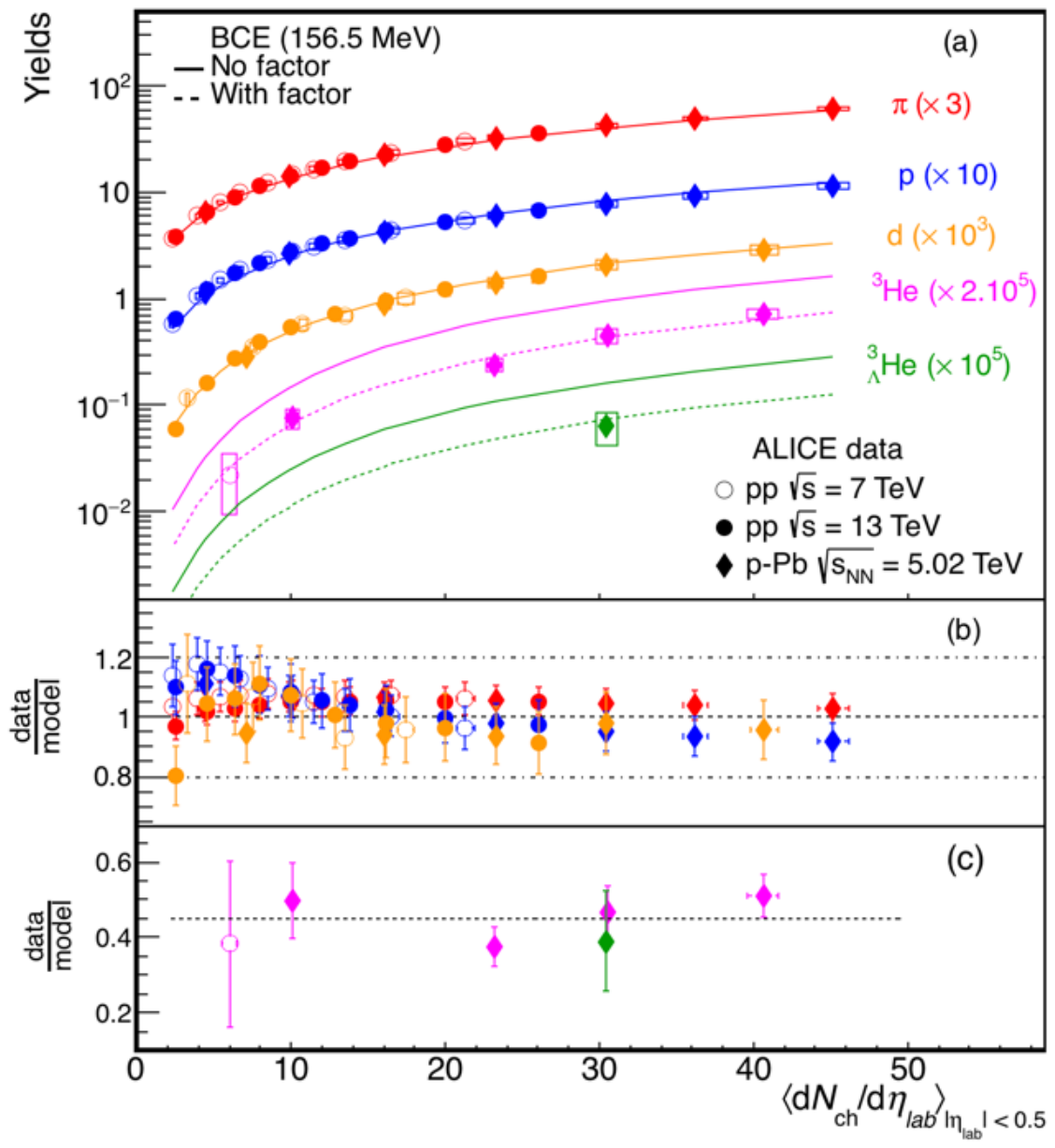} 
\caption{
(a) Yields of pions, protons and  light nuclei as a function of the charged particles multiplicities at mid-rapidity.
(b) Ratios of data over BCE thermal model for pions, protons, and deuterons. (c) Ratios for $^3$He and $^3_\Lambda$H. See text for more details.}
\label{fig:nucleiYields}
\end{figure}

Figure \ref{fig:nucleiYields} show the yields of pions, protons, deuterons, $^3$He and $ ^3_\Lambda$H as a function of $dN_{ch}/d\eta$ for pp collisions at $\sqrt{s}$ = 7 TeV and 13 TeV and p-Pb collisions at $\sqrt{s_{\rm NN}}$ = 5.02 TeV. The symbols are used to represent the experimental data measured by the ALICE collaboration~\cite{ALargeIonColliderExperiment:2021puh,ALICE:2020foi,ALICE:2019bnp,ALICE:2015wav,
ALICE:2017xrp,ALICE:2019fee}.
The solid lines represent yields of particles predicted from the thermal model that incorporates exact baryon conservation as given by  Eq. 7.  The light nuclei production in p-Pb collisions in various multiplicity bins are explained well by the thermal model.
The qualitative trend of $^3$He production as a function of multiplicity is nicely predicted by the BCE thermal model and the model results overpredict the 
observed yields. The middle and lower panel of the figure show the ratio of experimental yields over the BCE thermal  model predictions. 
The ratio in the lower panel is fitted with the polynomial function of zeroth order and the obtained parameter is 
$\lambda$ = 0.45 $\pm$ 0.03.
The dashed lines are the same predictions scaled by a factor $\lambda$ = 0.45 obtained from the linear fit in panel. 

Our analysis indicates that the yields of $^3$He and $^3_\Lambda$H show a noticeable trend of suppression as multiplicity decreases, suggesting the impact of the canonical effect. Furthermore, the quantitative representation of these yields deviates from the experimental data by a constant multiplicative factor which we call as a fugacity factor ($\lambda$) and can be interpreted as an effect of off-chemical equilibrium. 
This suggests that the particles with baryon number $>2$  deviate from the chemical equilibrium in pp and p-A collisions, with  "equilibrium factor" estimated to be around 0.45.   

\begin{figure}
\center
\includegraphics[scale=0.55]{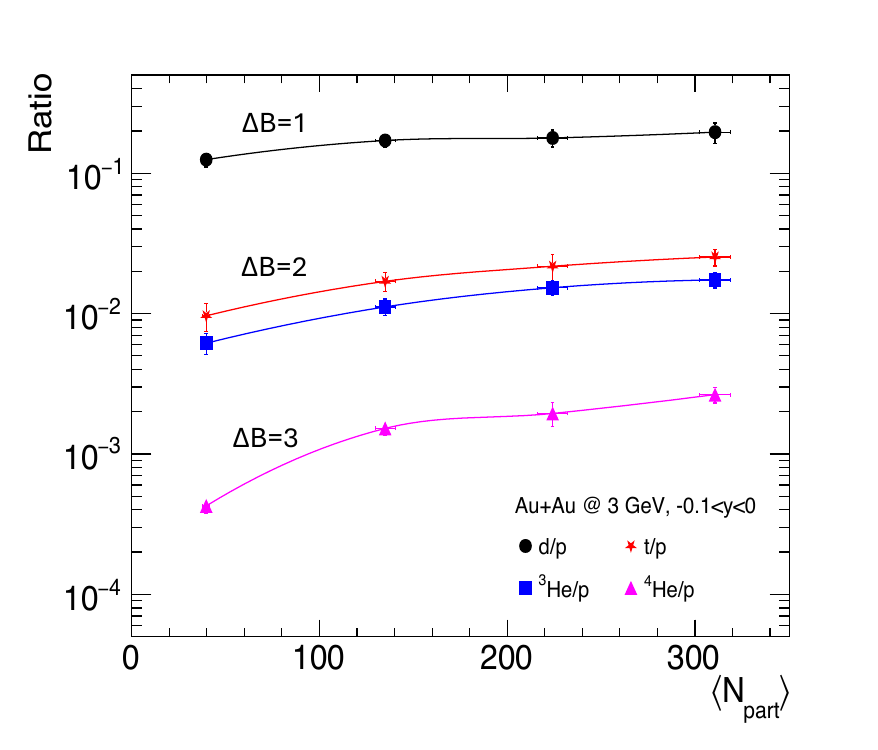} 
\vskip -0.4 cm
\caption{Light nuclei over protons ratios as a function of  the mean values of the number of participating nucleons ($<N_{\rm part}>$) measured in Au-Au collisions at $\sqrt{s_{\rm NN}}$ = 3 GeV by the STAR collaboration. $\Delta$B represents the number of baryons in numerator minus those in denominator. }
\label{fig:ratiotoP}
\end{figure}

Recently, the STAR collaboration has reported the systematic measurement of protons and light nuclei like deuterons, tritons, $^3$He, and $^4$He production in Au-Au collisions at $\sqrt{s_{\rm NN}}$ = 3 GeV~\cite{STAR:2023uxk}. 
Figure~\ref{fig:ratiotoP} shows the light nuclei over protons ratios as a function of  the mean values of the number of participating nucleons ($<N_{\rm part}>$). It is observed that the ratio is lower for higher value of 
$\Delta$B i.e. the difference between the number of baryons quantum number in numerator and denominator.
The 
light nuclei over protons ratios are decreasing as we go from central to peripheral collisions (i.e. with decreasing $<N_{\rm part}>$). The  suppression (slope) is 
more for higher mass and higher baryon quantum number. 
Thus the baryon suppression at 3 GeV may be interpreted as due to the baryon canonical effect. This needs further investigation.

\section*{Results and Discussions}
In conclusion, the experimental data show that the strangeness and baryon quantum numbers need to be treated canonically when discussing the particle production in the small colliding systems at the LHC energies and also in low energy collisions at the RHIC energies. 
The results underscore the importance of considering the canonical effect in small multiplicity and low energy collisions. The suppression of multi-strange and light nuclei particle yields in low multiplicity events as well as in low collision energy suggest that the production mechanisms are significantly influenced by the constraints of the canonical ensemble. This has implications on understanding the conditions under which multi-strange hadrons 
and light nuclei are produced in the high energy collisions.


Further investigation is needed to understand and explain the particle production mechanism  in the elementary and high energy heavy-ion  collisions.

\section*{Acknowledgments}
The author would like to thank Krzysztof Redlich, Peter Braun-Munzinger, Johanna Stachel, Pok Man Lo, and Lokesh Kumar for their support and collaboration. Special thanks to the ALICE and STAR collaborations for publishing data that enhances our understanding of physics.

\bibliographystyle{cas-model2-names}
\bibliography{main}

\end{document}